# On Performance of Peer Review for Academic Journals: Analysis Based on Distributed Parallel System

**Tan, Zong-Yuan**[1]    **Cai, Ning**[*1, 2]    **Zhou, Jian**[1, 3]    **Zhang, Sheng-Guo**[1, 2]

[1] College of Electrical Engineering, Northwest Minzu University, Lanzhou, China
[2] Key Laboratory of China's Ethnic Languages and Information Technology (Northwest Minzu University), Ministry of Education, Lanzhou, China
[3] College of Computer Science and Technology, Dalian University of Technology, Dalian, China

*Corresponding author: Cai, Ning (Email: caining91@tsinghua.org.cn)

This work is supported by National Natural Science Foundation (NNSF) of China (Grants 61867005 & 11805156), by Fundamental Research Funds for the Central Universities (Grants 31920160003, 31920170141 & 31920180115), by Program for Young Talents of State Ethnic Affairs Commission (SEAC) of China (Grant 2013-3-21), by Scientific Research Innovation Subject of Northwest Minzu University (Grant Yxm2018166), and by Research Project for Graduate Education and Teaching Reform of Northwest Minzu University.

**ABSTRACT** A simulation model based on parallel systems is established, aiming to explore the relation between the number of submissions and the overall standard of academic journals within a similar discipline under peer review. The model can effectively simulate the submission, review and acceptance behaviors of academic journals, in a distributed manner. According to the simulation experiments, it could possibly happen that the overall standard of academic journals may deteriorate due to excessive submissions.

## I. INTRODUCTION

Peer review is the cornerstone of science, whose essential purpose is to ensure that research publications are scientifically sound and are easily reusable [1-2]. As early as the 18$^{th}$ century, British Medical Association initially built peer review system for identifying the value of scientific literature more precisely [3]. At present, peer review has been already deemed as a crucial element by the vast majority of research institutions and scholars to evaluate the standard of academic journals [4-6]. For example, whether or not peer review is effectively conducted is a fundamental criterion for selecting scientific journals by some famous bibliographic databases, e.g. Scopus and EI Compendex. Though peer review has been challenged and criticized in academic community, it is still considered to be a gold criterion of scientific publication, particularly by playing an important role in filtering high standard manuscripts [7-8]. In addition, peer review could be viewed as a social process, and has also been applied in other activities such as fund allocation process for ensuring the fairness and effectiveness [9-10].

There already exist a number of works studying the mechanism of peer review mainly through modeling and simulations. Kovanis *et al.* [11] adopt an agent-based model (ABM) to empirically analyze the relation between peer review and publication systems. Based on [11], some innovative alternatives of peer review are evaluated to ameliorate the process of scientific publication in [12]. Similarly, Allesina [13] constructs a theoretical framework using an ABM to quantitatively study peer review systems. Thurner and Hanel [14] exploit a simple simulation model on peer review to study whether different types (correct, random and rational) of referees impact the standard of publication system. In [15], Righi and Takács evaluate the performance of peer review system by an ABM to select expectant high-quality scientific research for academic journals. Grimalda *et al.* [16-17] develop an ABM implemented by the Belief-Desire-Intention (BDI) platform to simulate peer review behaviors. Squazzoni and Gandelli [18-20] establish various simulation scenarios aiming to test how the performance of peer review is affected by the interactions of referees. Etkin [21] puts forward a new method and metric to assess the peer review process, aiming at assisting to enhance the standard of journals. Mrowinski *et al.* [22] create a directed weighted network regarding editorial workflows for researching the issue about review time. Moreover, for the sake of effectively performing peer review, Hak *et al.* [23] advise that journals should provide financial incentives to reward those referees who usually involve in review activities and create an R-index to quantify the contributions of those scientists as referees.

As the amount of academic manuscripts increases constantly, more load is imposed to the peer review system. Look and Sparks [24] point out that the charge of peer review paid for higher education in Britain is more than 110 million pounds every year. Peer review also costs much time and thereby the process of scientific publication is retarded. For instance, the review cycles in certain journals are about one month, whereas some reach up to one year or more. Since the limit of resources, the requirement for peer review due to the exponential growth of manuscripts could hardly be satisfied [25]. Review works would become more and more challenging, and further affect the sustainable development of scientific publication.

The primary aim of this paper is to explore how the overall standard of academic journals within a discipline is influenced by the number of submissions. For this end, a simple model is developed to simulate the submission, review and acceptance activities. Based on the model, the efficiency of peer review is analyzed experimentally. As the scale of submissions increases, a transition over three distinct phases is observable: 1) when the scale of submissions is minor, the trends of overall standard of journals ascend with the scale; 2) when the scale is appropriate, the trends of standard hold relatively steady; 3) when the scale is excessive, the trends would even descend, reversely.

The study is mainly based on the approach of social computing, which is rooted within the theoretical framework of parallel systems analysis [26-27]. Most of the parallel social systems are of large-scale and distributed, with some of the analytical properties potentially acquirable by certain theories in systems science, e.g. swarm stability [28-30], controllability [31-33], and systems synthesis [34-35]. Such a framework could be traced back to the dynamic model of segregation proposed by Schelling early in 1971 [36], who won the Nobel prize later, and also to the famous "sugarscape" model of Epstein and Axtell [37]. Now, it has been extensively and effectively applied for analyzing various complex social phenomena, e.g. [38]. By building and observing the behaviors of experimental systems, the objective of experiments conducted on parallel systems is not for comprehensively and quantitatively mimicking the real society, instead, it could be very conducive to revealing, validating, or enlightening the beingness of certain phenomena and principles, qualitatively.

Compared with the analogous studies in literature, e.g. [18-20], the work presented here holds three novel features. 1) The parallel model here is both quite simple and flexible, only grasping some most fundamental nature of peer review. For instance, the activities of authors and reviewers are all independent and the resource for reviewing is time. 2) Possible occurrence of a notable phenomenon is testified, which manifests that the overall standard of academic journals may deteriorate due to excessive submissions. 3) Based on the parallel model, comparison of the performance among different patterns of peer review can be conducted conveniently. To sum up, it is believable that the current work should be of benefit to apply, verify, and enrich relevant theories in two fields, namely systems analysis and bibliometrics.

The rest of this paper is organized as follows. Section 2 describes the framework of the model in detail. The relation between the overall standard of journals and the amount of submissions under peer review is analyzed in section 3. Finally, section 4 draws a brief conclusion.

## II. MODEL DESCRIPTIONS

A simple simulation model is constructed here, mainly for qualitatively analyzing the relation between the number of submissions and the overall standard of academic journals under peer review. It is worth mentioning that, for self-containment of the paper, the model is described in extenso. Some of the technical details are provided in this section merely to ensure reproducibility, but could be bypassed without influence to understanding the main idea.

The simulation model is discrete-timed, in which each iteration represents a synchronous round of publication for all journals, together with the submission, review, acceptance of manuscripts, and release of an issue by each journal behind it. The primary procedure of the model can be divided into several stages.

The first stage is the initialization of the model, where some fundamental information and parameter values are set, namely:

1) The number of journals.
2) The number of newly generated manuscripts per round $n$.
3) The number of submissions accepted by journals per round.
4) The number of referees.
5) Assign relevant parameter values, ($\alpha, \lambda, \beta, \gamma$), where $\alpha$ & $\lambda$ and $\beta$ & $\gamma$ jointly influence the quality estimation of manuscript by authors and the review by referees, respectively.
6) Define variable $q \in R$ to measure the quality of manuscript. Namely, the quality of each of the manuscript in our model can be scored by a value mostly falling within the scope of $[0,10]$, according to its quality.
7) Define quartile threshold $\theta$. Whether a manuscript is submitted to a journal in a quartile is determined by a quartile threshold. In other words, the quality of a submission should be higher than the quartile threshold of its target journal.

The second stage is the design of journal rankings. Analogous to the journal rankings of Journal Citation Report (JCR) and Chinese Academy of Sciences [39-40], the journals in our model are sorted and divided into quartiles, in which the first quartile is composed of journals with average article quality being the top 10%, whereas the second quartile 10%-25%, the third quartile 25%-50%, and the fourth quartile covers the remaining. Based on the simulation results, the journals can be sorted according to the average article quality, with the thresholds $\theta_1 \sim \theta_3$ naturally yielded, where $\theta_i$ is the average article quality of the lowest journal in $Q_i$ ($i = 1, 2, 3$), respectively.

The third stage is the creation of manuscripts. In reality, due to the fact that the quality of the majority of manuscripts is mainly mediocre leveled, those manuscripts with extremely high or low quality would be relatively less. In the current model, the quality $q_i$ ($i = 1, 2, \ldots$) of each manuscript follows positively skewed distribution with certain expectation and variance. The positively skewed distribution in the concrete program is implemented by gamma function, as shown in FIGURE 1.

**Remark 1.** It is known that data from observations which cannot have negative values are generally asymmetric with positive skew [41]. Records also reveal that the distribution of examination scores of students is often positively skewed [42]. The situation of examination scores in education is somewhat similar to manuscript qualities. Moreover, the limit of skewed distribution may tend to be Gaussian. Thus, it is reasonable to believe that the quality of manuscripts follows certain skewed distribution.

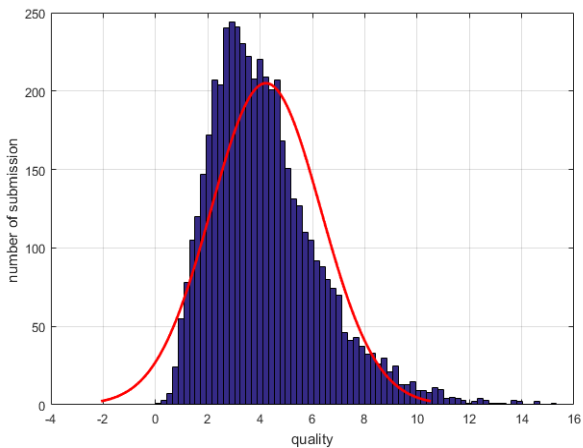

**FIGURE 1.** A sample of skewed quality distribution of manuscripts, with expectation being 4, variance being 1.062, and $n = 5000$. Red curve denotes gaussian distribution.

By observing the phenomenon from FIGURE. 1, it can be found that the number of manuscripts with high quality is rather minor, here high quality refers to $q > \theta_1$. This accords with the fact that the manuscripts with ultra-high quality are generally rare and their number holds relatively constant, despite the magnitude of overall submissions. Thus, as $n$ keeps on increasing, the number of ultra-high standard manuscripts would saturate, with only minor increase if above a limit. For ensuring that the manuscripts with high quality keep comparatively constant regardless of the variations of $n$, a numerical integration with respect to gamma function is performed in the model, which is to compute the area between the gamma distribution curve and the horizontal axis over the interval $[\theta_1, +\infty)$. Such an operation can be formulized as follows:

$$n_{q \geq \theta_1} = n \bullet \int_{\theta_1}^{+\infty} \Gamma(q) dq \quad (1)$$

where $n_{q \geq \theta_1}$ denotes the number of manuscripts with high quality; $\Gamma(q)$ denotes the gamma function.

The fourth stage is submission. Authors evaluate the quality of their manuscripts for selecting a suitable journal to submit. It is rational to assume that the estimation error depends on the academic ability of authors. Evidently, a professional scholar usually estimates his work more accurately, and vice versa, in most cases, a manuscript with high quality also implicates the creation of a competent author behind it. According to this principle, the following formula naturally yields:

$$\hat{q}_i = f(q_i) = q_i \bullet \xi_i \quad (i = 1, 2, \ldots, N) \quad (2)$$

where $q_i$ is the intrinsic quality of manuscript; $\hat{q}_i$ is the quality estimation by author; $N$ is the number of overall submissions per round, which is the sum of both $n$ and the previous rejected submissions. $\xi_i \in R$ reflects the estimation noise by author, which is a random value around 1, embodying the idiosyncrasy of individuals. The closer $\xi_i$ to 1, the more precise an estimation is. Thus, the variance of $\xi_i$ represents the magnitude of noise and should be negatively correlated to the manuscript quality. In the current settings, $\xi_i$ follows Gaussian distribution with expectation 1 and variance $\alpha / (q_i)^\lambda$, with parameters $\alpha$ and $\lambda$ being positive numbers that jointly determine the aggregation or dispersion of distribution curve, as illustrated in FIGURE 2. One can observe that in (2), $\hat{q} = [\hat{q}_1, \ldots, \hat{q}_N]^T = [f(q_1), \ldots, f(q_N)]^T$ satisfies $E(\hat{q}) = E(q)$. Therefore, $\hat{q}$ is unbiased estimator of $q$.

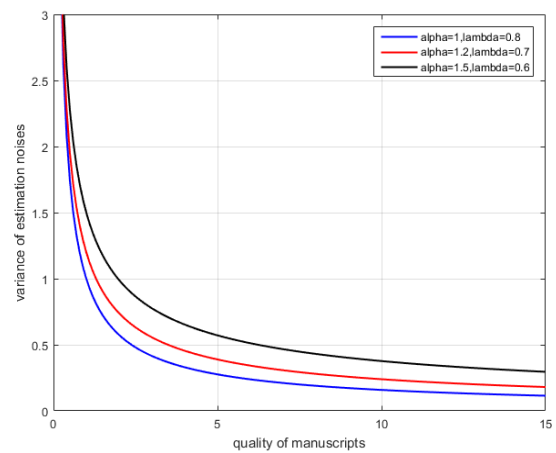

**FIGURE 2.** Change of variance for estimation noises with quality of manuscripts.

**Remark 2.** The overall shape of the curves is jointly determined by parameters $\alpha$ and $\lambda$. Anyway, the variance of estimation noises would gradually decline with the quality of manuscripts increasing.

The fifth stage is the peer review, which is the most important phase. Review is also an evaluation process to the quality of a manuscript. In the current model, two referees are randomly selected for every manuscript, with the number of manuscripts reviewed by each of them correspondingly augmented 1. Suppose that the accuracy of review scores is hinged on the number of manuscripts by referees. In reality, referees review manuscripts more carefully if they have sufficient time to do this, with the scores reported being relatively objective and precise; or the review would be rougher due to lack of time. According to this situation, the score can be calculated as follows in the model:

$$\begin{cases} \tilde{q}_i^{(1)} = g(q_i) = q_i \bullet \zeta_i^{(1)} \\ \tilde{q}_i^{(2)} = g(q_i) = q_i \bullet \zeta_i^{(2)} \end{cases} \quad (i = 1, 2, \ldots, N) \quad (3)$$

with $\tilde{q}_i^{(1)}$ and $\tilde{q}_i^{(2)}$ representing the review scores by both referees ; $\zeta_i^{(1)} \in R$ and $\zeta_i^{(2)} \in R$ reflecting the review noises, which are random values around 1, embodying the idiosyncrasy of individuals. The closer $\zeta_i^{(1)}$ or $\zeta_i^{(2)}$ to 1, the more precise a review is. Thus, the variances of $\zeta_i^{(1)}$ and $\zeta_i^{(2)}$ represent the magnitude of noises and should be positively correlated to the number of reviews imposed to referees. In the current settings, $\zeta_i^{(1)}$ and $\zeta_i^{(2)}$ are Gaussian random variables of expectation 1 and variance $\beta \bullet (k_i^{(1)})^\gamma$ & $\beta \bullet (k_i^{(2)})^\gamma$, where $k_i^{(1)}$ and $k_i^{(2)}$ represent the number of manuscripts reviewed by referees, and parameters $\beta$ and $\gamma$ are positive numbers that affect the review scores, as shown in FIGURE 3. Note that according to (3), $\tilde{q}^{(j)} = [\tilde{q}_1^{(j)}, \ldots, \tilde{q}_N^{(j)}]^T = [g(q_1^{(j)}), \ldots, g(q_N^{(j)})]^T$ ($j = 1, 2$) meets $E(\tilde{q}^{(j)}) = E(q)$. Consequently, $\tilde{q}^{(j)}$ are unbiased estimator of $q$.

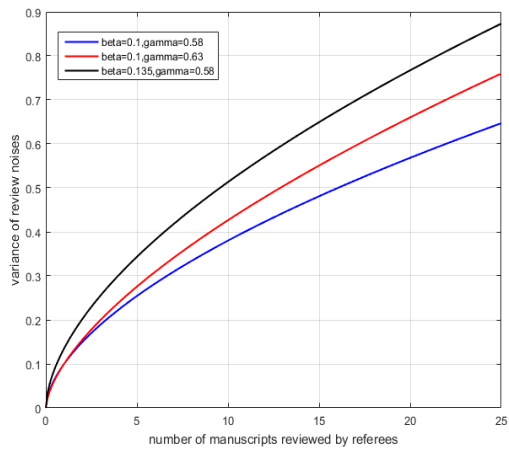

**FIGURE 3.** Change of variance of review noises with number of manuscripts reviewed by referees.

**Remark 3.** The overall shape of the curves is collectively influenced by parameters $\beta$ and $\gamma$. Anyway, as the number of manuscripts reviewed by referees increases persistently, the variance of review noises would also augment correspondingly.

The sixth stage is acceptance. The submissions to a journal are sorted according to the review scores. Each journal picks out a prescribed number of submissions with high scores. After that, the average quality of each journal can be computed, issue by issue. Those manuscripts rejected would be subsequently submitted to other journals in the next round.

The kernel of this model lies in two estimations to manuscript quality, namely the estimation by author and the estimation by reviewer. The noise of the first estimation is negatively correlated to the internal quality of a manuscript, while the noise of the second estimation is positively correlated to the review burden of a reviewer. Compared with relevant studies [18-20] in literature, such a layout is similar in mechanism but simpler in implementation, and it is possible to derive some analytical rules later.

By the end of this section, after a thorough description of a concrete model, one should note that the most distinct feature of the analysis based on parallel systems is that a parallel model stands independently and can itself be regarded as an instance of real systems, being a feasible alternative, implementation, reference, or compensation of the corresponding class of systems in reality. Therefore, it is not urgent to pursue accuracy in modeling, besides, such an approach is especially suitable for treating situations with ultra-complex mechanism or with unavailable data.

### III. SIMULATION RESULTS

This section will elucidate how the change trends of the overall standard of journals within a similar discipline are affected by the number of submissions under peer review. Admittedly, the primary purpose is to observe the variation of overall standard of journals, rather than to replicate any actual peer review system. It is worth mentioning that according to the experiments, the qualitative results are quite robust to parametric or even structural deviations. What illustrated later are just some typical samples.

The trends of average standard of journals in different quartiles can be effectively simulated, after relevant values are set, which are the number of manuscripts per round, the number of manuscripts with high quality and certain other parameters ($\alpha, \lambda, \beta, \gamma$), as illustrated in FIGURE. 4. It can be seen that if the number of submissions holds relatively scarce, the average standard of journals in the same quartile is fairly steady under peer review (although with some fluctuations). This may be due to that referees have sufficient time reviewing each submission and the review scores are more objective to reflect the intrinsic quality of submissions, so that journals can appropriately choose the high standard submissions.

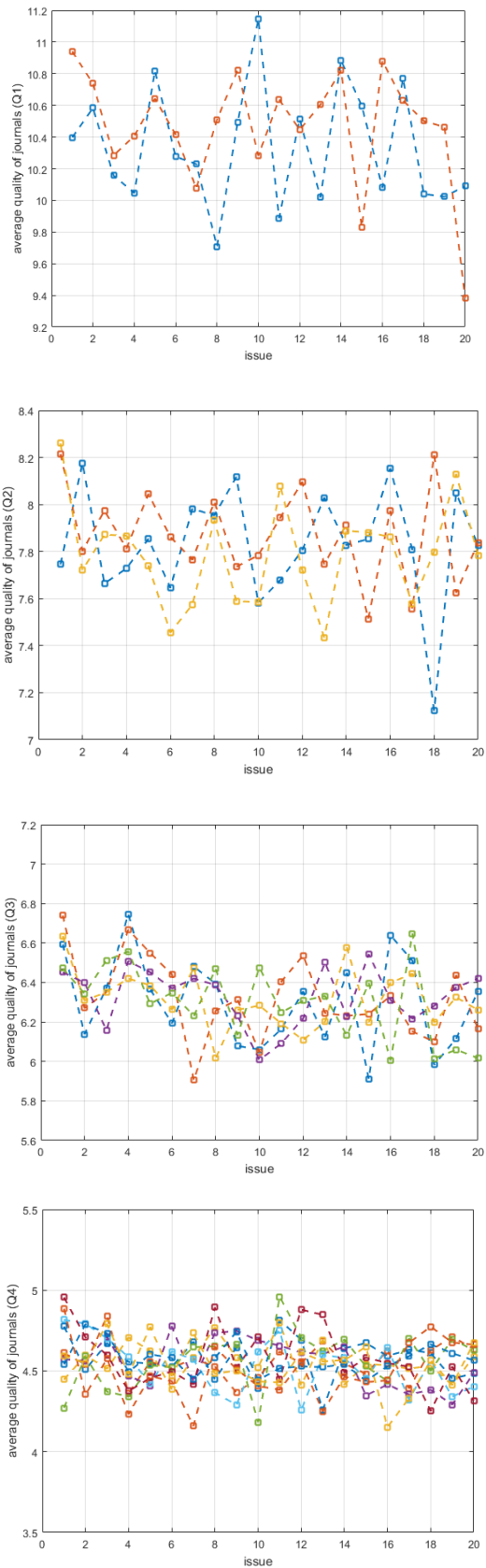

**FIGURE 4.** Variation of average quality of quartiles with issue, where $n=2000$, amount of submissions with high quality is 100 and $\alpha=1, \lambda=0.8, \beta=0.1, \gamma=0.58$.

As the submissions are increasing constantly, the precision of reviewer evaluation, i.e. the standard of review may descend. In order to analyze the situation, we simulate the relation between the variation trends of overall standard of journals and the number of submissions. See TABLEs Ⅰ, Ⅱ and Ⅲ.

TABLE Ⅰ
RELATION BETWEEN OVERALL STANDARD OF JOURNALS IN QUARTILES AND $n$

| No. | 400 | 600 | 800 | 1000 | 1200 | 1400 | 1600 | 1800 |
|---|---|---|---|---|---|---|---|---|
| Q1 | 10.25 | 10.21 | 10.23 | 10.26 | 10.22 | 10.25 | 10.32 | 10.4 |
| Q2 | 7.44 | 7.47 | 7.47 | 7.57 | 7.45 | 7.62 | 7.71 | 7.77 |
| Q3 | 5.79 | 5.79 | 5.74 | 5.72 | 5.84 | 6.04 | 6.14 | 6.23 |
| Q4 | 3.51 | 3.51 | 3.51 | 3.85 | 4.09 | 4.25 | 4.36 | 4.46 |

From the data shown in TABLE Ⅰ, on the condition that the number of submissions is minor, one can find that the overall level of journals keeps comparatively steady. The principal reason should be that the submissions are still scarce to meet the requirements. Therefore, the standard of journals retains rising with the number of submissions increasing. This indicates that sufficient submissions are beneficial to journals.

TABLE Ⅱ
RELATION BETWEEN OVERALL QUALITY OF JOURNALS IN QUARTILES AND $n$

| No. | 2000 | 2200 | 2400 | 2600 | 2800 | 3000 | 3200 | 3400 |
|---|---|---|---|---|---|---|---|---|
| Q1 | 10.34 | 10.29 | 10.23 | 10.19 | 10.14 | 10.07 | 10.08 | 9.99 |
| Q2 | 7.85 | 7.80 | 7.70 | 7.69 | 7.69 | 7.65 | 7.60 | 7.60 |
| Q3 | 6.30 | 6.34 | 6.35 | 6.29 | 6.26 | 6.23 | 6.20 | 6.16 |
| Q4 | 4.52 | 4.59 | 4.63 | 4.62 | 4.65 | 4.70 | 4.70 | 4.73 |

In comparison with TABLE Ⅰ, the efficiency of peer review would turn to reduce if the number of submissions persistently increases, as revealed in TABLE Ⅱ. One can see that the general standard of journals declines markedly, except the fourth quartile. When the number of submissions increases persistently, the tendency of overall standard of journals wholly descends, which is manifested in TABLE Ⅲ.

TABLE Ⅲ
RELATION BETWEEN OVERALL STANDARD OF JOURNALS IN QUARTILES AND $n$

| No. | 3600 | 3800 | 4000 | 4200 | 4400 | 4600 | 4800 | 5000 |
|---|---|---|---|---|---|---|---|---|
| Q1 | 10.03 | 10.08 | 9.96 | 9.95 | 9.83 | 9.94 | 9.88 | 9.84 |
| Q2 | 7.56 | 7.54 | 7.51 | 7.52 | 7.49 | 7.49 | 7.43 | 7.39 |
| Q3 | 6.12 | 6.11 | 6.07 | 6.06 | 6.01 | 6.01 | 5.98 | 5.94 |
| Q4 | 4.73 | 4.71 | 4.70 | 4.68 | 4.67 | 4.64 | 4.63 | 4.58 |

In sum, the overall standard of journals rises first, however a transition to dropping would appear later, with the submissions persistently increasing.

Next, when the number of submissions has saturated, the variation of overall standard of journals is exhibited in FIGURE. 5. One can clearly sense that although the average standard of journals in the first quartile has occasional undulation, the variation trends of it still remain comparatively steady generally. However in comparison, other quartiles bear evidently degressive tendency.

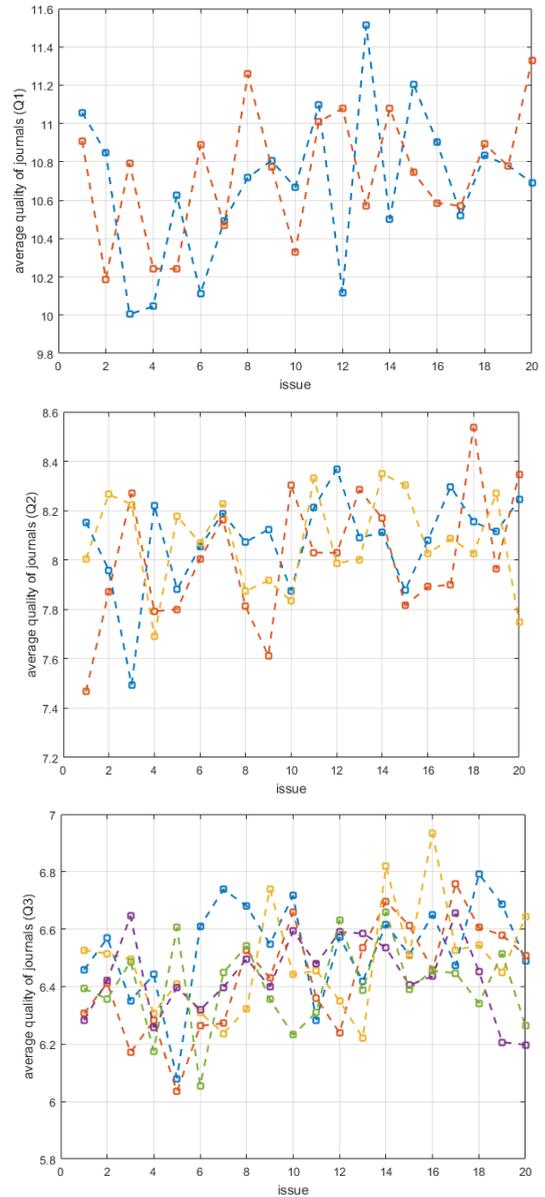

Additionally, consider the situation that if a manuscript is rejected by journals five times, then the manuscript would be abandoned. In accordance with the principle, the relation between the number of submissions and the overall standard of journals is further studied based on the current model. See FIGURE 6 and FIGURE 7.

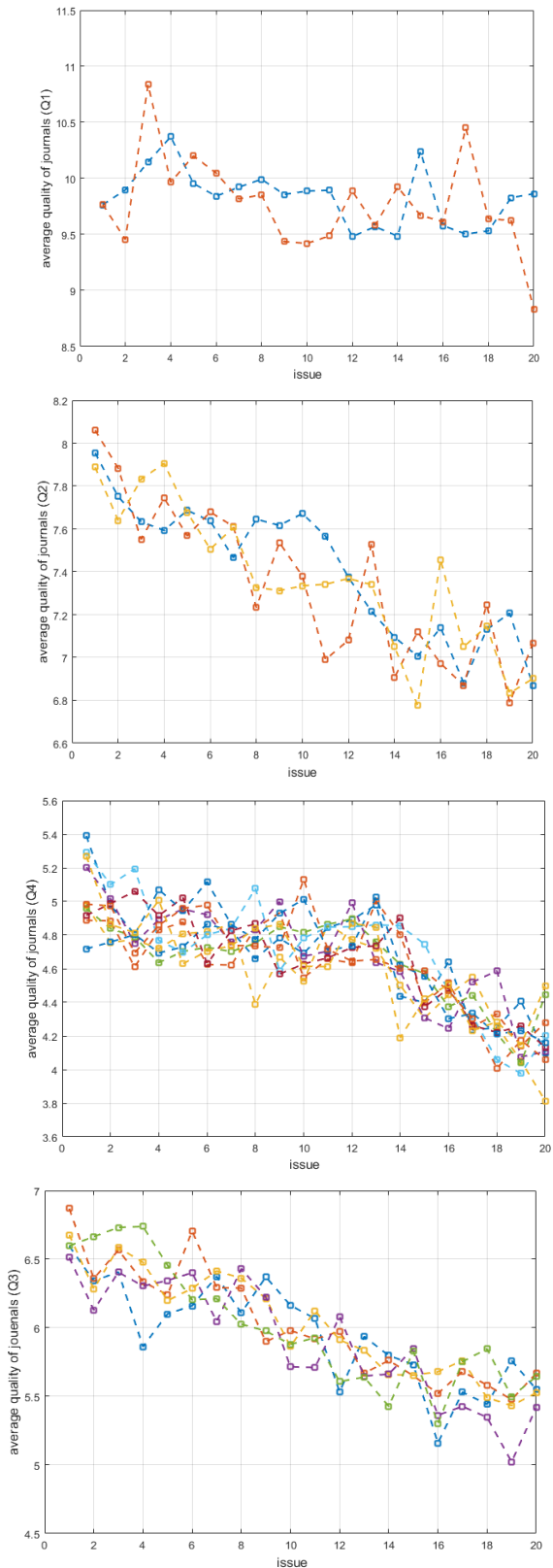

**FIGURE 5.** Variation of average quality of quartiles with issue, where $n = 5000$, amount of submissions with high quality is 110 and $\alpha = 1, \lambda = 0.8, \beta = 0.1, \gamma = 0.58$.

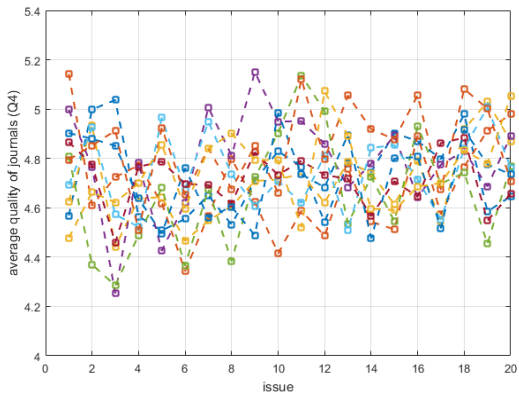

**FIGURE 6.** Variation of average quality of quartiles with issue, where $n<2500$, **amount of submissions with high quality is lower than 100 and** $\alpha=1, \lambda=0.8, \beta=0.1, \gamma=0.58$.

By observing the variation curves of FIGURE 6, it can be found that the overall trends of the average quality of articles in journals are roughly analogous. In other words, the average quality of journals is consistently ascending if the amount of submissions well accords the basic demand. It is evident that proper scale of submissions plays an essential role for preserving the overall standard of journals. Whereas by contrast, the trends of average standard of journals would rapidly descend if the amount of submissions increases significantly, then, they usually remain steady (although with a little ups and downs), as revealed in FIGURE 7.

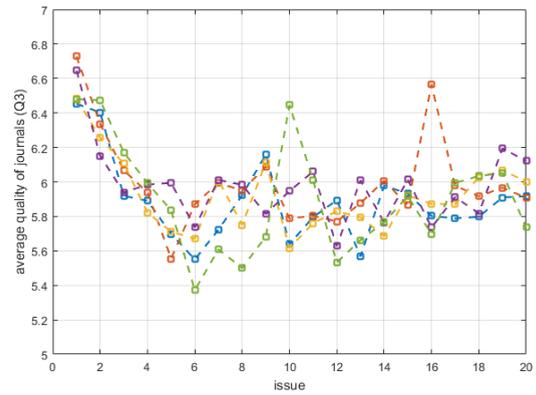

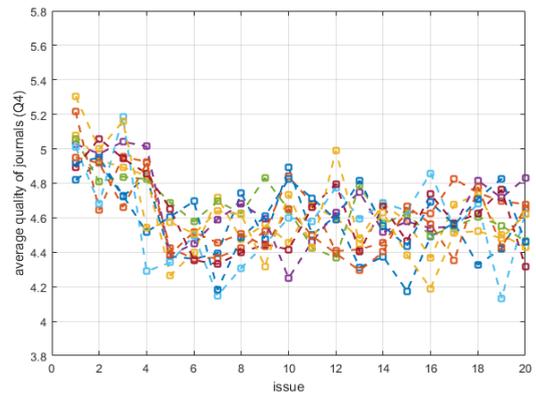

**FIGURE 7.** Variation of average quality of quartiles with issue, where $n>10000$, **amount of submissions with high quality is higher than 150 and** $\alpha=1, \lambda=0.8, \beta=0.1, \gamma=0.58$.

In practice, submission is not always right the first time. Thus, the statistics on the times of a submission rejected before acceptance is also conducted, as illustrated in FIGURE 8. One can clearly observe that as the amount of submissions increases, the times of submissions rejected before acceptance also correspondingly increase, whereas the number of submissions being right the first time markedly declines.

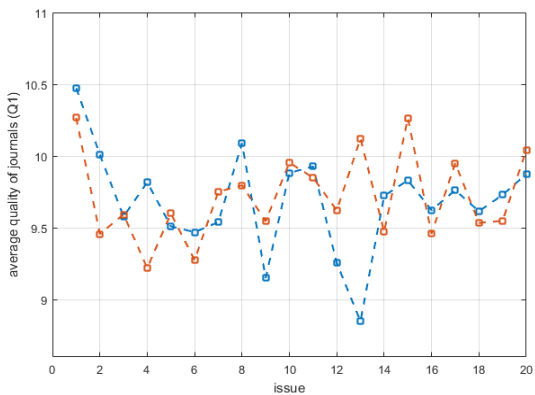

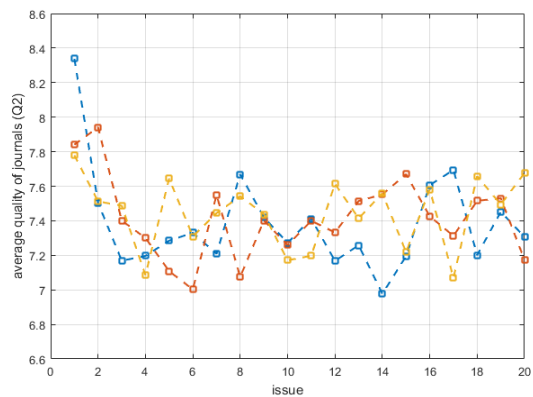

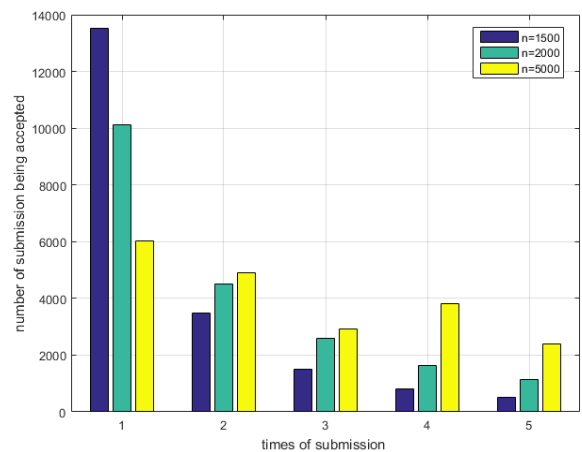

**FIGURE 8.** Times of submissions rejected before acceptance with different $n$.

## IV. CONCLUSION

Peer review is regarded as a gatekeeper of scientific publication, which helps to enhance the standard of academic journals. The primary purpose of this paper is to study the change trends of overall standard of academic journals with the amount of submissions increasing persistently under peer review. Our research approach is based on social computing, endeavoring to analyze the behaviors on academic journals through parallel simulations, including the behaviors like submission, review and acceptance, for evaluating the performance of peer review. The simulation results indicate that the trends of overall standard of academic journals remain relatively steady or rise persistently if the amount of submissions is appropriate; while as the amount keeps on increasing to a saturation state, the trends would decline. Generally speaking, it could possibly happen that the overall standard of journals may deteriorate due to excessive submissions. One should be aware that intuition taken for granted is virtually undependable in science, until it is verified by sufficient evidence and rational inference. Without the approach of social computing applied, then there might be only diffusive speculations, due to the lack of sufficient data. This implies the key significance of social computing based on experimental study of parallel systems. Our endeavor here is a typical practice in this way and should be of benefit to provide theoretical support for real-world applications. Evidently, some future work can be further continued. For instance, based on the current model, certain novel patterns of peer review could be studied for promoting the overall standard of academic journals; meanwhile, relevant analytical rules might also be revealed by deeper delving.

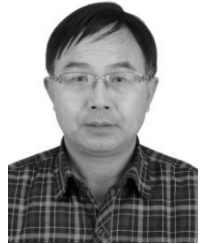

**Sheng-Guo Zhang** received his M.S. degree in mechatronic engineering in 2007 and Ph.D. degree in manufacturing engineering and automation in 2012 from the Department of Precision Instruments and Mechanology, Tsinghua University, Beijing, China.

He is currently a professor with the Key Lab of China's National Languages Information Technology and also with the School of Electrical Engineering, Northwest Minzu University, Lanzhou 730030, China. His research interests include mechatronic system technology, microcomputer control technology, manufacturing automation technology, and electrical automation technology. Email: zhangshengguo@tsinghua.org.cn

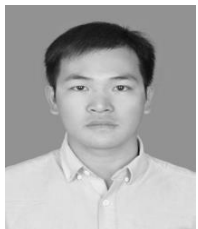

**Zong-Yuan Tan** received the B.E. degree in Information Engineering from Guilin University of Technology in 2016. He is currently a graduate student in Northwest Minzu University. His research interest includes distributed computing and informetrics.

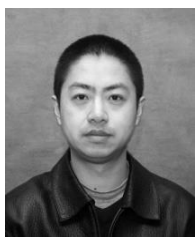

**Ning Cai** received the B.E. degree in Control Engineering from Tsinghua University in 1996, the M.E. degree in Computer Science from Chinese Academy of Sciences in 1999, and the Ph.D. degree in Control Theory from Tsinghua University in 2010. He was shortly with Tianjin Polytechnic University and Tsinghua University from 2010 to 2011. Since 2011, he has been an associate professor with Northwest Minzu University. His current research interest focuses on systems simulation and analysis based on dynamical complex networks.

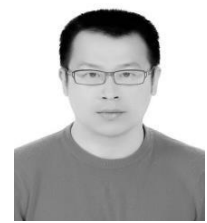

**Jian Zhou** received the B.E. degree in Electronic Engineering from the College of Technology and Engineering, Lanzhou University of Technology in 2013, the M.E. degree in Computer Science from the College of Electrical Engineering, Northwest Minzu University in 2018. He is currently pursuing his Ph.D. degree in the College of Computer Science and Technology, Dalian University of Technology. His research interest includes systems simulation, informetrics and machine learning.